\documentclass[nofootinbib,preprint]{revtex4}%
\usepackage{hyperref}
\usepackage{amsmath,amssymb}
\usepackage{graphicx}
\usepackage{epstopdf}
\usepackage{amsmath}
\usepackage{amsfonts}
\usepackage{amssymb}%
\providecommand{\U}[1]{\protect\rule{.1in}{.1in}}
\begin{document}
\title{A Refined Holographic QCD Model and QCD Phase Structure}
\author{Yi Yang}
\email{yiyang@mail.nctu.edu.tw}
\affiliation{Department of Electrophysics, National Chiao Tung University}
\affiliation{National Center for Theoretical Science, Hsinchu, ROC}
\author{Pei-Hung Yuan}
\email{phy.pro.phy@gmail.com}
\affiliation{Institute of Physics, National Chiao Tung University, Hsinchu, ROC}
\date{\today}

\begin{abstract}
We consider the Einstein-Maxwell-dilaton system with an arbitrary kinetic
gauge function and a dilaton potential. A family of analytic solutions is
obtained by the potential reconstruction method. We then study its holographic
dual QCD model. The kinetic gauge function can be fixed by requesting the
linear Regge spectrum of mesons. We calculate the free energy to obtain the
phase diagram of the holographic QCD model.

\end{abstract}
\maketitle
\tableofcontents

%

\setcounter{equation}{0}
\renewcommand{\theequation}{\arabic{section}.\arabic{equation}}%

\section{Introduction}

To study phase structure of QCD is a challenging and important task. It is
well known that QCD is in the confinement and chiral symmetry breaking phase
at the low temperature and small chemical potential, while it is in the
deconfinement and chiral symmetry restored phase at the high temperature and
large chemical potential. Thus it is widely believed that there exists a phase
transition between these two phases. To obtain the phase transition boundary
in the $T-\mu$ phase diagram is a rather difficult task because the QCD
coupling constant becomes very strong near the phase transition region and the
conventional perturbative method does not work well. Moreover, with the
nonzero physical quark masses presented, part of the phase transition line
will weaken to a crossover for a range of temperature and chemical potential
that makes the phase structure of QCD more complicated to study. Locating the
critical point where the phase transition converts to a crossover is an
important but difficult task. For a long time, the technique of lattice QCD is
the only reliable method to attack these problems. Although lattice QCD works
very well for zero density, it encounters the sign problem when considering
finite density or chemical potential, i.e. $\mu\neq0$. However, the most
interesting region in the QCD phase diagram is at finite density. The most
concerned subjects, such as heavy-ion collisions and compact stars in
astrophysics, are all related to QCD at finite density. Recently, lattice QCD
has developed some techniques to solve the sign problem, such as reweighting
method, imaginary chemical potential method and the method of expansion in
$\mu/T$. Nevertheless, these techniques are only able to deal with the cases
of small chemical potentials and quickly lost control for the larger chemical
potential. See \cite{1009.4089} for a review of the current status of lattice QCD.

On the other hand, using the recently developed idea of AdS/CFT correspondence
from string theory, one is able to study QCD in the strongly coupled region by
studying its weakly coupled dual gravitational theory, the so called
holographic QCD. The models which are directly constructed from string theory
are called top-down models. The most popular top-down models are D3-D7
\cite{0306018,0311270,0304032,0611099} model and D4-D8 (Sakai-Sugimoto) model
\cite{0412141,0507073}. In these top-down holographic QCD models, confinement
and chiral symmetry phase transitions in QCD have been addressed and been
translated into geometric transformations in the dual gravity theories. Meson
spectrums and their decay constants have also been calculated and compared
with the experimental data with surprisingly consistency. Although the
top-down QCD models describe many important properties in realistic QCD, the
meson spectrums obtained from those models can not realize the linear Regge
trajectories. To solve this problem, another type of holographic models have
been developed, i.e. bottom-up models, such as the hard wall model
\cite{0501128} and the later refined soft-wall model \cite{0602229}. In the
original soft-wall model, the IR correction of the dilaton field was put by
hand to obtain the linear Regge behavior of the meson spectrum. However, since
the fields configuration is put by hand, it does not satisfy the equations of
motion. To get a fields configuration which is both consistent with the
equation of motions and realizes the linear Regge trajectory, dynamical
soft-wall models were constructed by introduce a dilaton potential
\cite{0801.4383,0806.3830} consistently. On the other hand, the
Einstein-dilaton and Einstein-Maxwell-dilaton models have been widely studied
numerically \cite{0804.0434,1006.5461,1012.1864,1108.2029,1201.0820} to
investigate the thermodynamical properties and explore the phase structure in
QCD. Recently, by a potential reconstruction method, analytic solutions can be
obtained in the Einstein-dilaton model \cite{1103.5389} and similarly in the
Einstein-Maxwell-dilaton model \cite{1201.0820,1209.4512}.

In this paper, we try to combine the techniques of the dynamical soft-wall
model and the potential reconstruction methods to study QCD phase diagram as
well as the linear Regge spectrum of mesons. We consider a
Einstein-Maxwell-dilaton system with an arbitrary kinetic gauge function and a
dilaton potential as in \cite{1301.0385}. A family of analytic solutions are
obtained by the potential reconstruction method. We then study its holographic
dual QCD model. The kinetic gauge function can be fixed by requesting the
meson spectrums satisfy the linear Regge trajectories. By studying the
thermodynamics of the Einstein-Maxwell-dilaton background, we calculate the
free energy to obtain the phase diagram of our holographic QCD model. We
compute the different equation of states in our model and discuss their behaviors.

The paper is organized as follows. In section II, we consider the
Einstein-Maxwell-dilaton system with a dilaton potential as well as a gauge
kinetic function. By potential reconstruction method, we obtain a family of
analytic solutions with arbitrary gauge kinetic function and warped factor. We
then fix the gauge kinetic function by requesting the meson spectrums to
realize the linear Regge trajectories. By choosing a proper warped factor, we
obtain the final form of our analytic solution. In section III, we study the
thermodynamics of our gravitational background and compute the free energy to
get the phase diagram. We conclude our result in section IV.%

\setcounter{equation}{0}
\renewcommand{\theequation}{\arabic{section}.\arabic{equation}}%

\section{Einstein-Maxwell-Dilaton System}

We consider a 5-dimensional Einstein-Maxwell-dilaton system with probe flavor
fields as in \cite{1301.0385}. The action of the system have two parts, the
background part and the matter part,%
\begin{equation}
S=S_{b}+S_{m}.
\end{equation}
The background action includes a gravity field $g_{\mu\nu}$, a Maxwell field
$A_{\mu}$ and a neutral dilatonic scalar field $\phi$. While the matter action
includes two flavor fields $\left(  {A}_{\mu}^{L},{A}_{\mu}^{R}\right)  $,
representing the left-handed and right-handed gauge fields, respectively. The
Kaluza-Klein modes of these 5d flavor gauge fields describe the degrees of
freedom of mesons on the 4d boundary. We will treat the matter fields as probe
fields and do not consider their backreaction to the background.

In Einstein frame, the background action and the matter action can be written
as%
\begin{align}
S_{b}  &  =\dfrac{1}{16\pi G_{5}}\int d^{5}x\sqrt{-g}\left[  {R-\frac{f\left(
\phi\right)  }{4}F^{2}}-\dfrac{1}{2}\partial_{\mu}\phi\partial^{\mu}%
\phi-V\left(  \phi\right)  \right]  ,\label{action-b}\\
S_{m}  &  =-\dfrac{1}{16\pi G_{5}}\int d^{5}x\sqrt{-g}{\frac{f\left(
\phi\right)  }{4}}\left(  {F_{V}^{2}+F_{\tilde{V}}^{2}}\right)  .
\label{action-m}%
\end{align}
where we have expressed the flavor fields\ ${A}^{L}$ and ${A}^{R}$\ in terms
of the vector meson and pseudovector meson fields $V$ and $\tilde{V}$,%
\begin{equation}
{A}^{L}=V+\tilde{V}\text{, \ \ }{A}^{R}=V-\tilde{V}.
\end{equation}
The equations of motion can be derived from the actions (\ref{action-b}) and
(\ref{action-m}) as%
\begin{align}
\nabla^{2}\phi &  =\frac{\partial V}{\partial\phi}+\frac{1}{4}\frac{\partial
f}{\partial\phi}\left(  F^{2}+{F_{V}^{2}+F_{\tilde{V}}^{2}}\right)
,\label{eom1}\\
\nabla_{\mu}\left[  f(\phi)F^{\mu\nu}\right]   &  ={{0,}}\\
\nabla_{\mu}\left[  f(\phi)F_{V}^{\mu\nu}\right]   &  ={{0,}}\label{eom3}\\
\nabla_{\mu}\left[  f(\phi)F_{\tilde{V}}^{\mu\nu}\right]   &  ={0,}\\
R_{\mu\nu}-\frac{1}{2}g_{\mu\nu}R  &  =\frac{f(\phi)}{2}\left(  F_{\mu\rho
}F_{\nu}^{\rho}-\frac{1}{4}g_{\mu\nu}F^{2}+\left\{  {F_{V},F_{\tilde{V}}%
}\right\}  \right)  +\frac{1}{2}\left[  \partial_{\mu}\phi\partial_{\nu}%
\phi-\frac{1}{2}g_{\mu\nu}\left(  \partial\phi\right)  ^{2}-g_{\mu\nu
}V\right]  . \label{eom5}%
\end{align}
First, we will solve the gravitational background in the above
Einstein-Maxwell-dilaton system. We consider the following ansatz for the
metric, the Maxwell field and the dilaton field%
\begin{align}
ds^{2}  &  =\dfrac{L^{2}e^{2A\left(  z\right)  }}{z^{2}}\left[  -g(z)dt^{2}%
+\frac{dz^{2}}{g(z)}+d\vec{x}^{2}\right]  ,\label{metric}\\
\phi &  =\phi\left(  z\right)  \text{, \ \ }A_{\mu}=A_{t}\left(  z\right)  ,
\label{ansatz}%
\end{align}
where $z=0$ corresponds to the conformal boundary of the 5d spacetime and we
will set the radial $L$ of $AdS_{5}$ space to be unit in the following of this
paper. By turning off the probe fields $V$ and $\tilde{V}$ in the equations of
motion (\ref{eom1}-\ref{eom5}), the equations of motion for the background
fields become%
\begin{align}
\phi^{\prime\prime}+\left(  \frac{g^{\prime}}{g}+3A^{\prime}-\dfrac{3}%
{z}\right)  \phi^{\prime}+\left(  \frac{z^{2}e^{-2A}A_{t}^{\prime2}f_{\phi}%
}{2g}-\frac{e^{2A}V_{\phi}}{z^{2}g}\right)   &  =0,\label{eom-phi}\\
A_{t}^{\prime\prime}+\left(  \frac{f^{\prime}}{f}+A^{\prime}-\dfrac{1}%
{z}\right)  A_{t}^{\prime}  &  =0,\label{eom-At}\\
A^{\prime\prime}-A^{\prime2}+\dfrac{2}{z}A^{\prime}+\dfrac{\phi^{\prime2}}{6}
&  =0,\label{eom-A}\\
g^{\prime\prime}+\left(  3A^{\prime}-\dfrac{3}{z}\right)  g^{\prime}%
-e^{-2A}z^{2}fA_{t}^{\prime2}  &  =0,\label{eom-g}\\
A^{\prime\prime}+3A^{\prime2}+\left(  \dfrac{3g^{\prime}}{2g}-\dfrac{6}%
{z}\right)  A^{\prime}-\dfrac{1}{z}\left(  \dfrac{3g^{\prime}}{2g}-\dfrac
{4}{z}\right)  +\dfrac{g^{\prime\prime}}{6g}+\frac{e^{2A}V}{3z^{2}g}  &  =0.
\label{eom-V}%
\end{align}
We impose the regular boundary conditions at the horizon $z=z_{H}$ and the
asymptotic AdS condition at the boundary $z\rightarrow0$ as follows,%
\begin{align}
A_{t}\left(  z_{H}\right)   &  =g\left(  z_{H}\right)  =0,\label{bc-zH}\\
A\left(  0\right)   &  =-\sqrt{\dfrac{1}{6}}\phi\left(  0\right)  \text{,
\ }g\left(  0\right)  =1,\label{bc-0}\\
A_{t}\left(  0\right)   &  =\mu+\rho z^{2}+\cdots,
\end{align}
where $\mu$ is quark chemical potential and $\rho$ is quark density. By the
potential reconstruction method, the above equations of motion (\ref{eom-phi}%
-\ref{eom-V}) can be analytically solved as%
\begin{align}
\phi^{\prime}\left(  z\right)   &  =\sqrt{-6\left(  A^{\prime\prime}%
-A^{\prime2}+\dfrac{2}{z}A^{\prime}\right)  },\label{phip}\\
A_{t}\left(  z\right)   &  =\sqrt{\dfrac{-1}{\int_{0}^{z_{H}}y^{3}%
e^{-3A}dy\int_{y_{g}}^{y}\dfrac{x}{e^{A}f}dx}}\int_{z_{H}}^{z}\dfrac{y}%
{e^{A}f}dy,\label{At}\\
g\left(  z\right)   &  =1-\dfrac{\int_{0}^{z}y^{3}e^{-3A}dy\int_{y_{g}}%
^{y}\dfrac{x}{e^{A}f}dx}{\int_{0}^{z_{H}}y^{3}e^{-3A}dy\int_{y_{g}}^{y}%
\dfrac{x}{e^{A}f}dx},\label{gp}\\
V\left(  z\right)   &  =-3z^{2}ge^{-2A}\left[  A^{\prime\prime}+3A^{\prime
2}+\left(  \dfrac{3g^{\prime}}{2g}-\dfrac{6}{z}\right)  A^{\prime}-\dfrac
{1}{z}\left(  \dfrac{3g^{\prime}}{2g}-\dfrac{4}{z}\right)  +\dfrac
{g^{\prime\prime}}{6g}\right]  , \label{V}%
\end{align}
where the gauge kinetic function $f\left(  z\right)  $ and the warped factor
$A\left(  z\right)  $ are two arbitrary functions. Different choices of the
functions $f\left(  z\right)  $ and $A\left(  z\right)  $ will give different
physically consistent backgrounds. The undetermined integration constant
$y_{g}$ in the above solution is related to the chemical potential $\mu$ of
the dual QCD as%
\begin{equation}
\mu=-\sqrt{\dfrac{-1}{\int_{0}^{z_{H}}y^{3}e^{-3A}dy\int_{y_{g}}^{y}\dfrac
{x}{e^{A}f}dx}}\int_{0}^{z_{H}}\dfrac{y}{e^{A}f}dy, \label{mu}%
\end{equation}
in which $y_{g}$ can be solved in term of the chemical potential $\mu$ once
the manifest forms of the gauge kinetic function $f\left(  z\right)  $ and the
warped factor $A\left(  z\right)  $ are given.

We next consider the 5d probe vector field $V$ whose equation of motion has
been derived in (\ref{eom3}),%
\begin{equation}
\nabla_{\mu}\left[  f\left(  \phi\right)  F_{V}^{\mu\nu}\right]  ={{0.}}%
\end{equation}
With the gauge $V_{z}=0$, the equation of motion of the transverse vector
field $V_{\mu}$ $\left(  \partial^{\mu}V_{\mu}=0\right)  $ in the above
gravitational background becomes%
\begin{equation}
\dfrac{1}{g}\nabla^{2}V+V^{\prime\prime}+\left(  \dfrac{g^{\prime}}{g}%
+\dfrac{f^{\prime}}{f}+A^{\prime}-\dfrac{1}{z}\right)  V^{\prime
}=0,\label{eqV}%
\end{equation}
where the prime is\ the derivative respect to $z$. By expanding the vector
field $V$ for discrete values of 4d momentum $k_{n}=\left(  \omega_{n},\vec
{p}_{n}\right)  $,%
\begin{equation}
V\left(  x,z\right)  =\sum_{k_{n}}e^{ik_{n}\cdot x}X\psi_{n}\left(  z\right)
\text{, \ \ }X=\left(  \dfrac{z}{e^{A}fg}\right)  ^{1/2}%
\end{equation}
we bring the equation of motion (\ref{eqV}) into the form of the
Schr\"{o}dinger equation%
\begin{equation}
-\psi_{n}^{\prime\prime}+U\left(  z\right)  \psi_{n}=m_{n}^{2}\left(
z\right)  \psi_{n},
\end{equation}
with the potential function and the "energy dependent" mass as%
\begin{equation}
U\left(  z\right)  =\dfrac{2X^{\prime2}}{X^{2}}-\dfrac{X^{\prime\prime}}%
{X}\text{, \ \ }m_{n}\left(  z\right)  =\sqrt{\dfrac{\omega_{n}^{2}}%
{g^{2}\left(  z\right)  }-\dfrac{\vec{p}_{n}^{2}}{g\left(  z\right)  }}%
\end{equation}
In the limit of zero chemical potential and zero temperature, i.e. $g\left(
z\right)  \equiv1$, we expect that the discrete spectrum of the vector mesons
obeys the linear Regge trajectories. In this case, the above Schr\"{o}dinger
equation reduces to%
\begin{equation}
-\psi_{n}^{\prime\prime}+U\left(  z\right)  \psi_{n}=m_{n}^{2}\psi
_{n},\label{Shordinger}%
\end{equation}
where $m_{n}^{2}=\omega_{n}^{2}-\vec{p}_{n}^{2}$. Following \cite{0602229},
the simple choice of $f\left(  z\right)  =e^{\pm cz^{2}-A\left(  z\right)  }$
brings the potential to the form%
\begin{equation}
U\left(  z\right)  =\dfrac{3}{4z^{2}}+c^{2}z^{2}.\label{potential}%
\end{equation}
The Schr\"{o}dinger equations (\ref{Shordinger}) with the above potential
(\ref{potential}) have the discrete eigenvalues
\begin{equation}
m_{n}^{2}=4cn,\label{mass}%
\end{equation}
which is linear in the energy level $n$ as we expect for the vector spectrum
at zero temperature and zero density.

Once we fixed the gauge kinetic function $f\left(  z\right)  =e^{\pm
cz^{2}-A\left(  z\right)  }$, the Eq.(\ref{mu}) can be solved to get the
integration constant $y_{g}$ in term of the chemical potential $\mu$
explicitly as%
\begin{equation}
e^{cy_{g}^{2}}=\dfrac{\int_{0}^{z_{H}}y^{3}e^{-3A}e^{cy^{2}}dy}{\int
_{0}^{z_{H}}y^{3}e^{-3A}dy}+\dfrac{\left(  1-e^{cz_{H}^{2}}\right)  ^{2}%
}{2c\mu^{2}\int_{0}^{z_{H}}y^{3}e^{-3A}dy}. \label{yg}%
\end{equation}
Put the integration constant $y_{g}$ back into the solution (\ref{phip}%
-\ref{V}),\ we finally write down our solution as%
\begin{align}
\phi^{\prime}\left(  z\right)   &  =\sqrt{-6\left(  A^{\prime\prime}%
-A^{\prime2}+\dfrac{2}{z}A^{\prime}\right)  },\label{phip-A}\\
A_{t}\left(  z\right)   &  =\mu\dfrac{e^{cz^{2}}-e^{cz_{H}^{2}}}%
{1-e^{cz_{H}^{2}}},\label{At-A}\\
g\left(  z\right)   &  =1+\dfrac{1}{\int_{0}^{z_{H}}y^{3}e^{-3A}dy}\left[
\dfrac{2c\mu^{2}}{\left(  1-e^{cz_{H}^{2}}\right)  ^{2}}\left\vert
\begin{array}
[c]{cc}%
\int_{0}^{z_{H}}y^{3}e^{-3A}dy & \int_{0}^{z_{H}}y^{3}e^{-3A}e^{cy^{2}}dy\\
\int_{z_{H}}^{z}y^{3}e^{-3A}dy & \int_{z_{H}}^{z}y^{3}e^{-3A}e^{cy^{2}}dy
\end{array}
\right\vert -\int_{0}^{z}y^{3}e^{-3A}dy\right]  ,\\
V\left(  z\right)   &  =-3z^{2}ge^{-2A}\left[  A^{\prime\prime}+3A^{\prime
2}+\left(  \dfrac{3g^{\prime}}{2g}-\dfrac{6}{z}\right)  A^{\prime}-\dfrac
{1}{z}\left(  \dfrac{3g^{\prime}}{2g}-\dfrac{4}{z}\right)  +\dfrac
{g^{\prime\prime}}{6g}\right]  . \label{V-A}%
\end{align}
Note that our final solution (\ref{phip-A}-\ref{V-A}) depends on the warped
factor $A\left(  z\right)  $. The choice of $A\left(  z\right)  $ is arbitrary
provided it satisfies the boundary condition (\ref{bc-0}).%

\setcounter{equation}{0}
\renewcommand{\theequation}{\arabic{section}.\arabic{equation}}%

\section{Phase Structure}

In \cite{1301.0385}, a simple form of the warped factor has been studied,%
\begin{equation}
A\left(  z\right)  =-\dfrac{c}{3}z^{2}-bz^{4}. \label{A}%
\end{equation}
The parameters $c\simeq1.16GeV^{2}$ and $b\simeq0.273GeV^{4}$ were determined
by fitting the lowest two quarkonium states, $m_{J/\psi}=3.096GeV$ and
$m_{\psi^{\prime}}=3.685GeV$, as well as comparing the phase transition
temperature at $\mu=0$ to the lattice QCD simulation of $T_{HP}\simeq0.6GeV$
in \cite{1111.4953}. With these parameters, the authors of \cite{1301.0385}
argued that the system is to describe the heavy quarks with the deconfinement
phase transition. However, in this work, we will consider another parameter
regime of $b$ and $c$ to study the light quarks with the chiral symmetry
breaking phase transition.

\subsection{Meson Spectrum}

We consider the same form (\ref{A}) of the warped factor $A\left(  z\right)  $
as in \cite{1301.0385}. We will determine the parameter $c$ by fitting the
meson spectrum (\ref{mass}) to the experimental data. Instead of fitting the
quarkonium states made up of heavy quarks in \cite{1301.0385}, we now consider
mesons made up of light quarks, i.e. $\rho$ meson and its excitations. We take
the experimental data of the lowest six excitations of $\rho$ meson from
PDG2007 \cite{PDG2007}. From the date, we fit the parameter $c\simeq
0.227GeV^{2}$ in the mass formula (\ref{mass}) by using the standard $\chi
^{2}$ fit \cite{0710.0988,0804.2731}. The experiment data and our fitting are
list in Table \ref{data}.

\begin{table}[ptb]
\begin{tabular}
[c]{|c|c|c|c|c|c|c|}\hline
$n$ & $1$ & $2$ & $3$ & $4$ & $5$ & $6$\\\hline
$m_{\rho^{n}}$ & $0.77$ & $1.45$ & $1.70$ & $1.90$ & $2.15$ & $2.27$\\\hline
Fitting & $0.95$ & $1.35$ & $1.65$ & $1.90$ & $2.13$ & $2.34$\\\hline
\end{tabular}
\caption{The experiment data (in GeV) of $\rho$ meson and its excitations from
PDG2007 \cite{PDG2007}.}
\label{data}
\end{table}

\subsection{Black Hole Thermodynamics}

Using the black hole solution we obtained in the previous section,%
\begin{equation}
ds^{2}=\dfrac{e^{2A\left(  z\right)  }}{z^{2}}\left[  -g(z)dt^{2}+\frac
{dz^{2}}{g(z)}+d\vec{x}^{2}\right]  ,
\end{equation}
it is easy to calculate the Hawking-Bekenstein entropy%
\begin{equation}
s=\dfrac{e^{3A\left(  z_{H}\right)  }}{4z_{H}^{3}}, \label{entropy}%
\end{equation}
and the Hawking temperature%
\begin{equation}
T=\dfrac{z_{H}^{3}e^{-3A\left(  z_{H}\right)  }}{4\pi\int_{0}^{z_{H}}%
y^{3}e^{-3A\left(  y\right)  }dy}\left[  1-\dfrac{2c\mu^{2}\left(
e^{cz_{H}^{2}}\int_{0}^{z_{H}}y^{3}e^{-3A\left(  y\right)  }dy-\int_{0}%
^{z_{H}}y^{3}e^{-3A\left(  y\right)  }e^{cy^{2}}dy\right)  }{\left(
1-e^{cz_{H}^{2}}\right)  ^{2}}\right]  . \label{temperature}%
\end{equation}
To continue, we need to fix the parameter $b$ in\ the warped factor (\ref{A})
for our black hole background.

\begin{figure}[h]
\begin{center}
\includegraphics[
height=2in,
width=3in
]{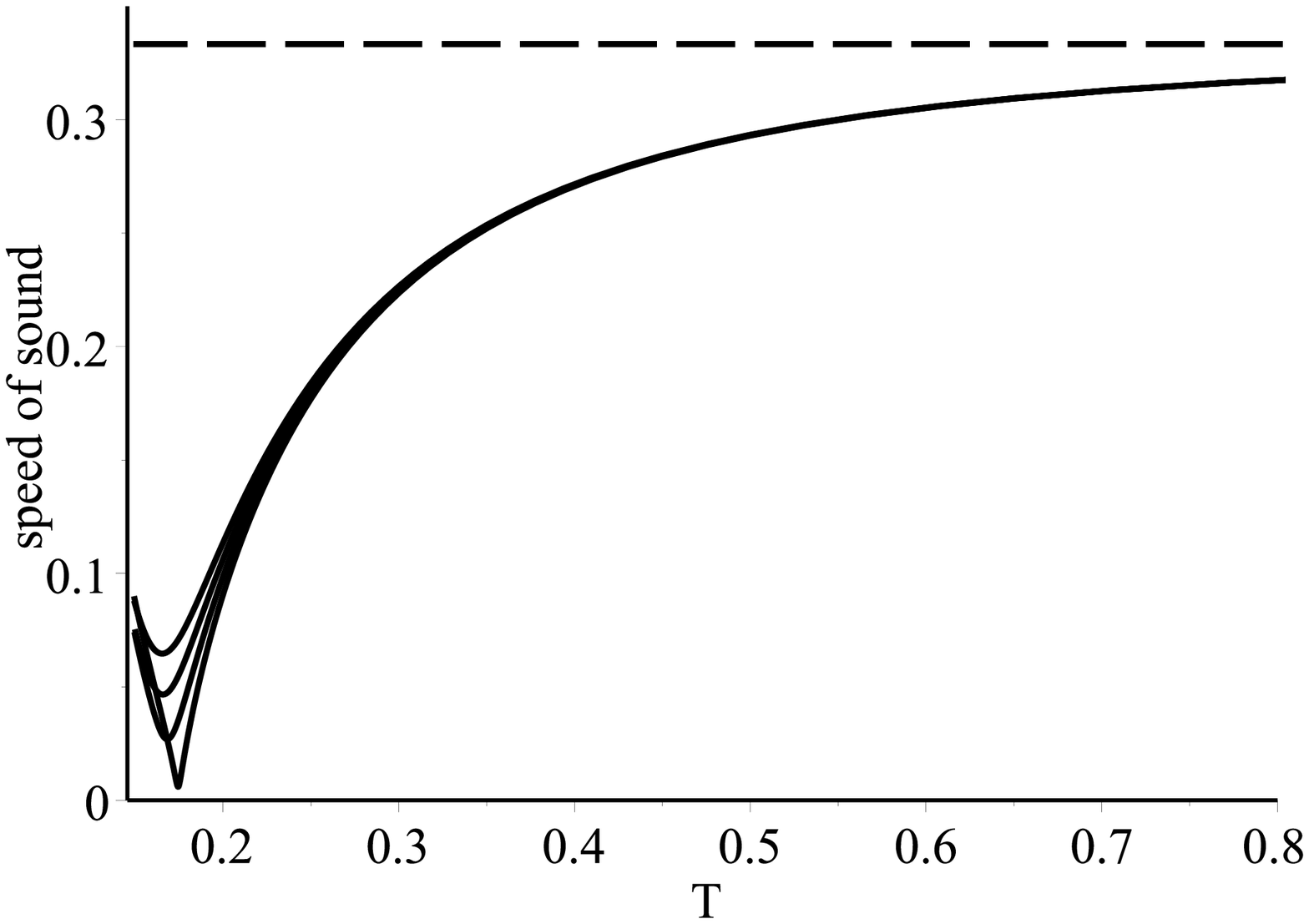}\hspace*{0.5cm} \includegraphics[
height=2in,
width=2.9in
]{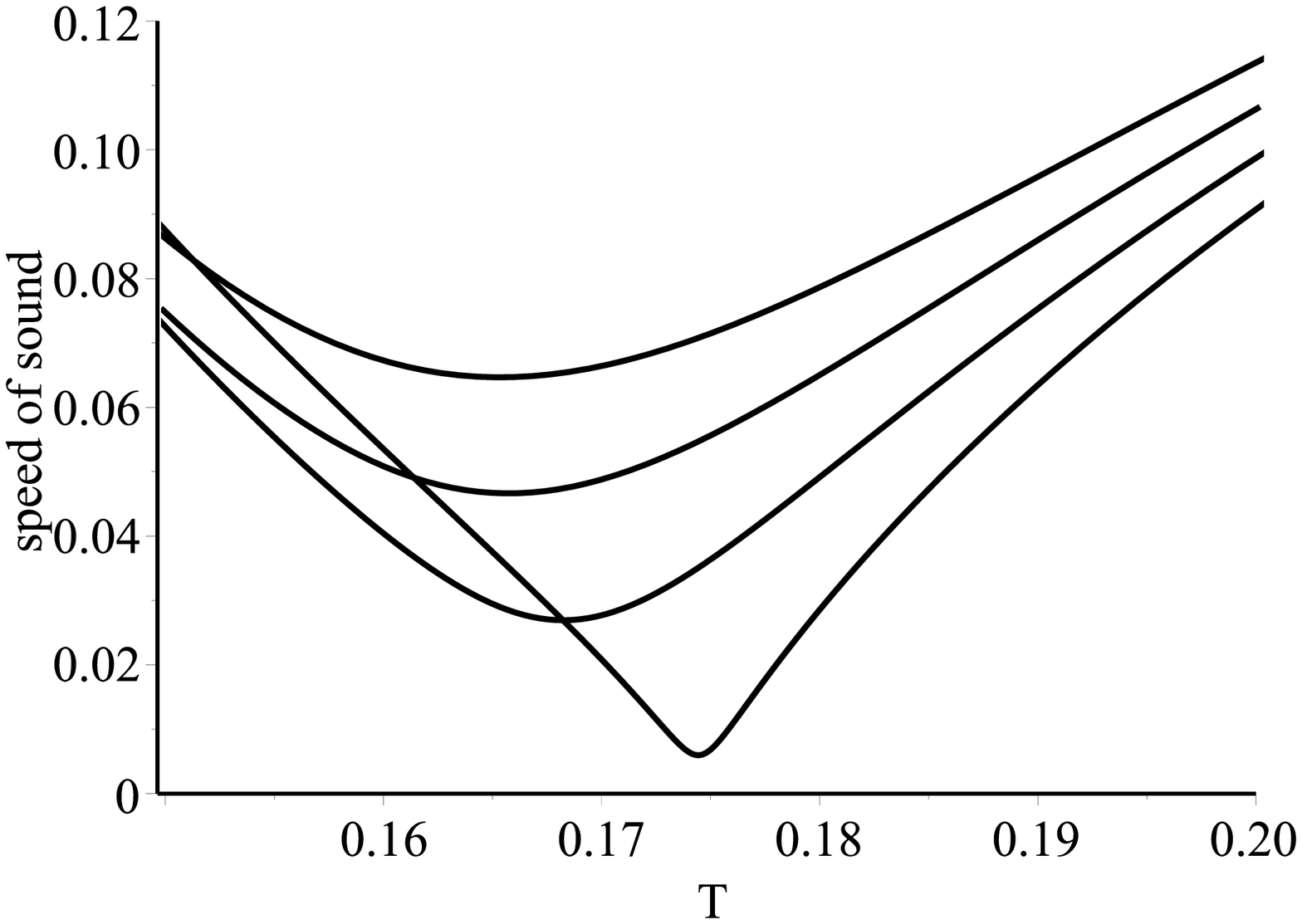}\vskip -0.05cm \hskip 0.15 cm \textbf{( a ) } \hskip 7.5 cm
\textbf{( b )}
\end{center}
\caption{The squared speed of sound vx
.s. temperature at $\mu=0$ for different
values of parameter $b$. Curves from top to bottom correspond to
$b=-0.0005,-0.0007,-0.0009,-0.0011GeV^{4}$. We enlarge a rectangle region in
(a) into (b) to see the detailed structure. For different values of the
parameter $b$, the corresponding quasi-transition temperature $T_{0}$, i.e.
the position of the minimum value,\ changes.}%
\label{csb}%
\end{figure}

We will fix the parameter $b$ by fitting the phase transition temperature
$T_{0}$\ at the vanishing chemical potential obtained from lattice QCD. It is
well known that, for QCD with quark mass, the phase transition becomes a
crossover at low chemical potential. There is no realizing order parameter to
describe a crossover. Nevertheless, we can define a quasi-transition
temperature by looking at a rapid change for certain observable. In
\cite{1403.1179}, the authors argued that the quasi-transition temperature for
a crossover is not uniquely defined and therefore depends on the observable
used to define it. Basically any observable that exhibits a non-differentiable
behavior at the critical temperature can be used to define the
quasi-transition temperature at a crossover. It is not surprised that the
quasi-transition temperature changes with the observable used to define it. In
this case, people use the transition region \cite{1005.3508}, in which
different observable may have their characteristic points at different
temperature values. Since the temperature dependencies of the various
observable play a more crucial role than any single quasi-transition
temperature value, we will use the speed of sound to define the
quasi-transition temperature in this paper. The speed of sound is defined as%
\begin{equation}
c_{s}^{2}=\dfrac{\partial\ln T}{\partial\ln s}.
\end{equation}
At $\mu=0$, the Hawking temperature (\ref{temperature}) reduces to%
\begin{equation}
T\left(  z_{H}\right)  =\dfrac{z_{H}^{3}e^{-3A\left(  z_{H}\right)  }}%
{4\pi\int_{0}^{z_{H}}y^{3}e^{-3A\left(  y\right)  }dy},
\end{equation}
and the speed of sound becomes%
\[
c_{s}^{2}=\dfrac{z_{H}^{4}e^{-3A\left(  z_{H}\right)  }}{3\left[
1-z_{H}A^{\prime}\left(  z_{H}\right)  \right]  \int_{0}^{z_{H}}%
y^{3}e^{-3A\left(  y\right)  }dy}-1.
\]
In FIG. \ref{csb}, we plot the squared speed of sound v.s. temperature at
vanishing chemical potential for several values of parameter $b$ in the warped
factor (\ref{A}). We can see that, for each curve, there is a rapid change at
a temperature around $150\sim200MeV$, which we can use to define the
quasi-transition temperature $T_{0}$\ of the crossover at $\mu=0$. For
different values of the parameter $b$, quasi-transition temperature $T_{0}$,
i.e. the position of the minimum value,\ changes as shown in FIG. \ref{csb}.
By taking the commonly used value $T_{0}\approx170MeV$, we can fix the
parameter as $b=-6.25\times{10}^{-4}GeV^{4}$.

\subsection{Phase Diagram}

\begin{figure}[h]
\begin{center}
\includegraphics[
height=2in,
width=3in
]{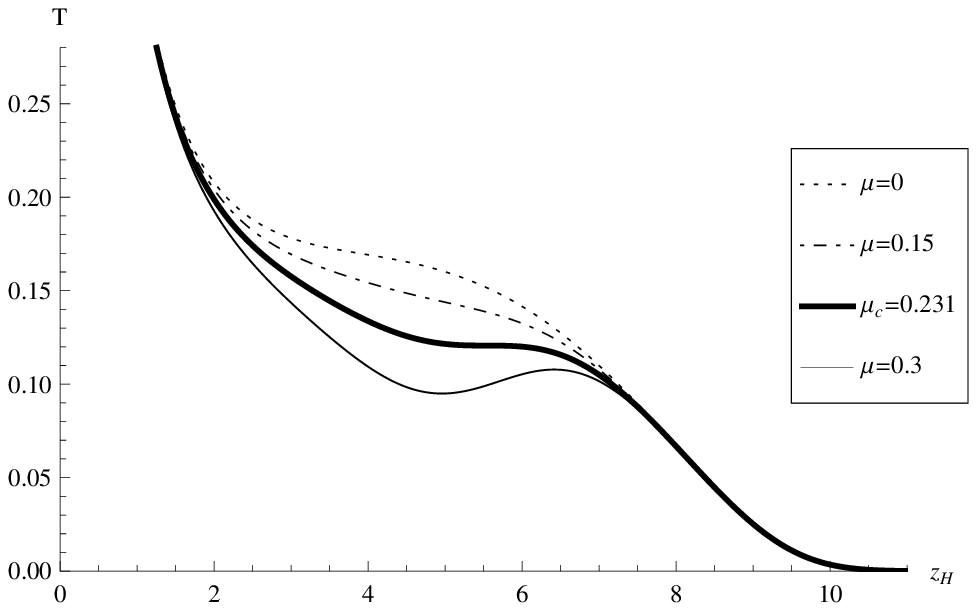}\hspace*{0.5cm} \includegraphics[
height=2in,
width=2.9in
]{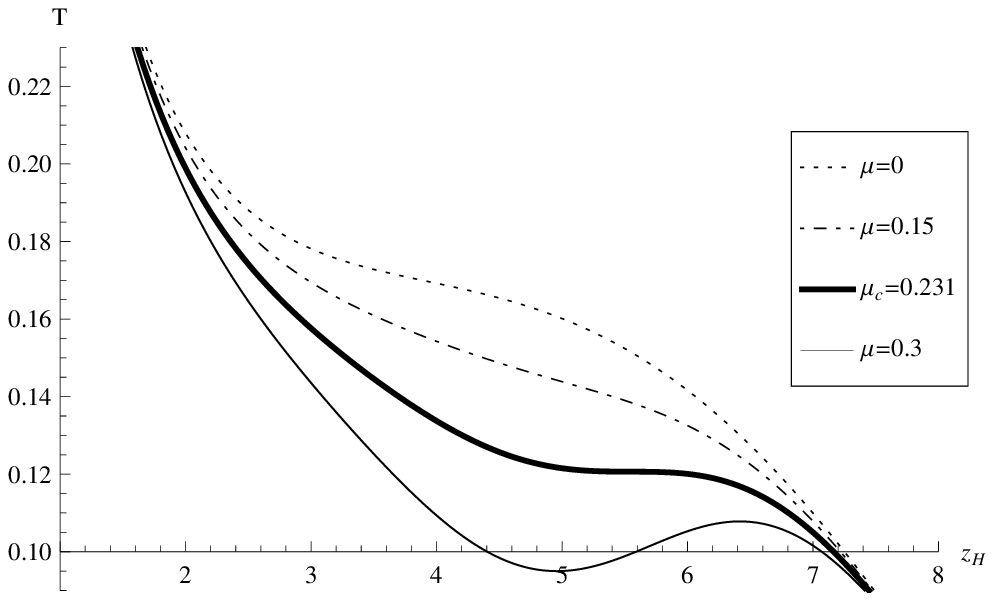}\vskip -0.05cm \hskip 0.15 cm \textbf{( a ) } \hskip 7.5 cm \textbf{(
b )}
\end{center}
\caption{The temperature v.s. horizon at different chemical potentials
$\mu=0,0.15,0.231,0.3GeV$. We enlarge a rectangle region in (a) into (b) to
see the detailed structure. For $0<\mu<\mu_{c}$, the temperature decreases
monotonously to zero; while for $\mu>\mu_{c}$, the temperature has a local
minimum. At $\mu_{c}\simeq0.231GeV$, the local minimum reduces to a inflection
point.}%
\label{T}%
\end{figure}

For different chemical potentials, the temperature dependence on the horizon
$z_{H}$ is showed in FIG. \ref{T}. For vanishing or small chemical potential
$0\leq\mu\leq\mu_{c}=0.23148GeV$, the temperature decreases monotonously to
zero; while for $\mu>\mu_{c}$, the temperature bends up and goes down again to
zero. Therefore, for certain range, the same temperature corresponds to three
different horizons as indicated in (b) of FIG. \ref{T}. This temperature
behavior implicates that a phase transition happens at certain temperature for
$\mu>\mu_{c}$.

\begin{figure}[h]
\begin{center}
\includegraphics[
height=2in,
width=3in
]{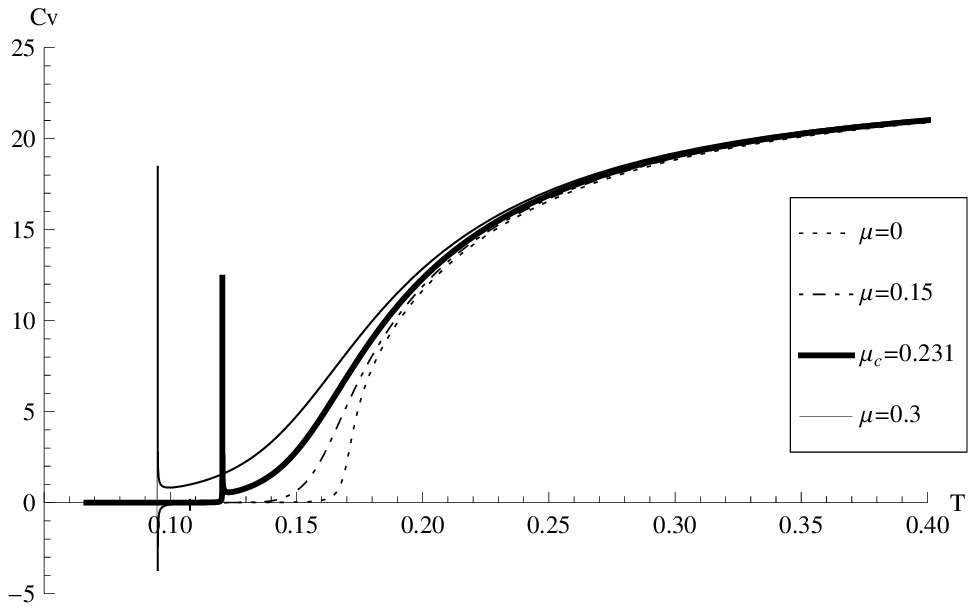}\hspace*{0.5cm} \includegraphics[
height=2in,
width=2.9in
]{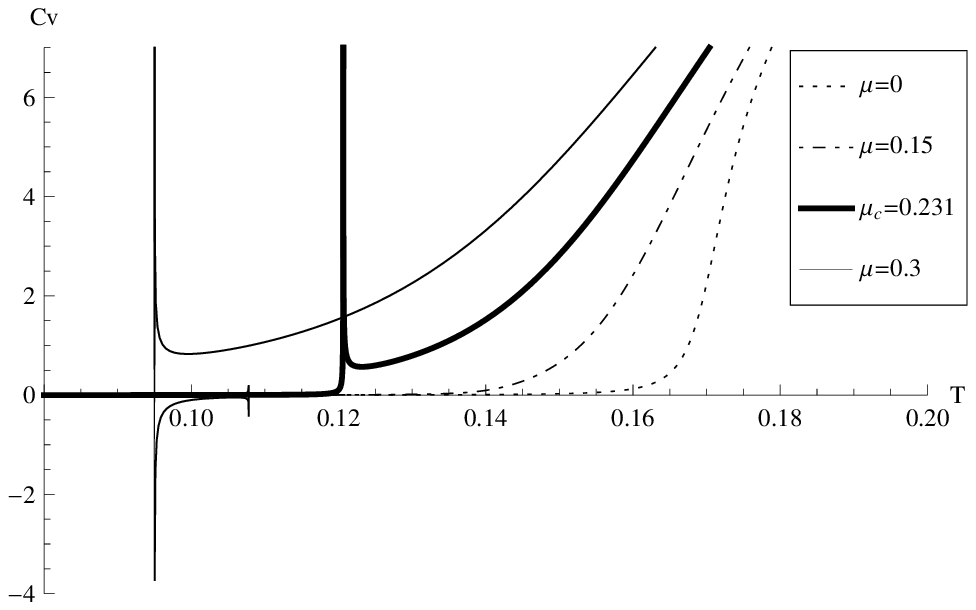}\vskip -0.05cm \hskip 0.15 cm \textbf{( a ) } \hskip 7.5 cm
\textbf{( b )}
\end{center}
\caption{The specific heat v.s. temperature at different chemical potentials
$\mu=0,0.15,0.231,0.3GeV$. We enlarge a rectangle region in (a) into (b) to
see the detailed structure. For $0\leq\mu\leq\mu_{c}$, the specific heat is
always positive, $C_{V}>0$ implies that the black hole with any temperature is
thermodynamically stable. While for $\mu>\mu_{c}$, $C_{V}$ could be negative
for a range of $T$ where the black hole is thermodynamically unstable.}%
\label{C}%
\end{figure}

To determine the thermodynamically stability, we plot specific heat $C_{V}$
v.s. temperature $T$ in FIG. \ref{C}, where the specific heat $C_{V}$ is
defined as%
\begin{equation}
C_{V}=T\left(  \dfrac{\partial s}{\partial T}\right)  _{\mu}.
\end{equation}

In the $C_{V}-T$ diagram, the negative value of the specific heat corresponds
to the thermodynamically instability. For $0\leq\mu\leq\mu_{c}$, the specific
heat is always positive. $C_{V}>0$ implies that the black hole with any
temperature is thermodynamically stable. While for $\mu>\mu_{c}$, $C_{V}$
could be negative for a range of $T$ where the black hole is thermodynamically
unstable. Thus one of the three horizons corresponding to the same temperature
is thermodynamically unstable and the black hole would never take that state.
However, there still left two horizons which are both thermodynamically stable
and are possible realistic states. To determined which one is physically
preferred out of the two thermodynamically stable states, we need to compare
their free energies.

\begin{figure}[h]
\begin{center}
\includegraphics[
height=2in,
width=3in
]{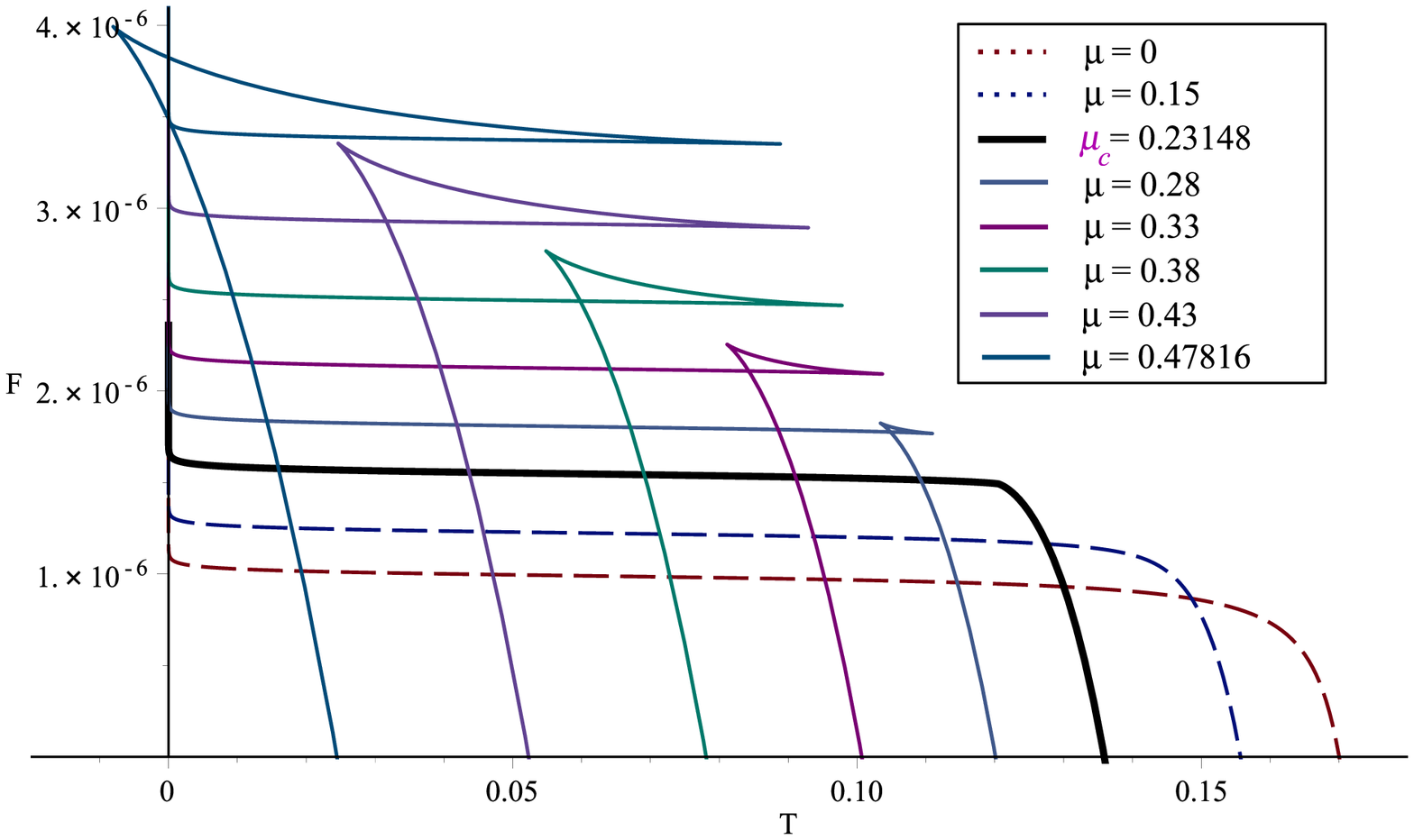}\hspace*{0.5cm} \includegraphics[
height=2in,
width=3in
]{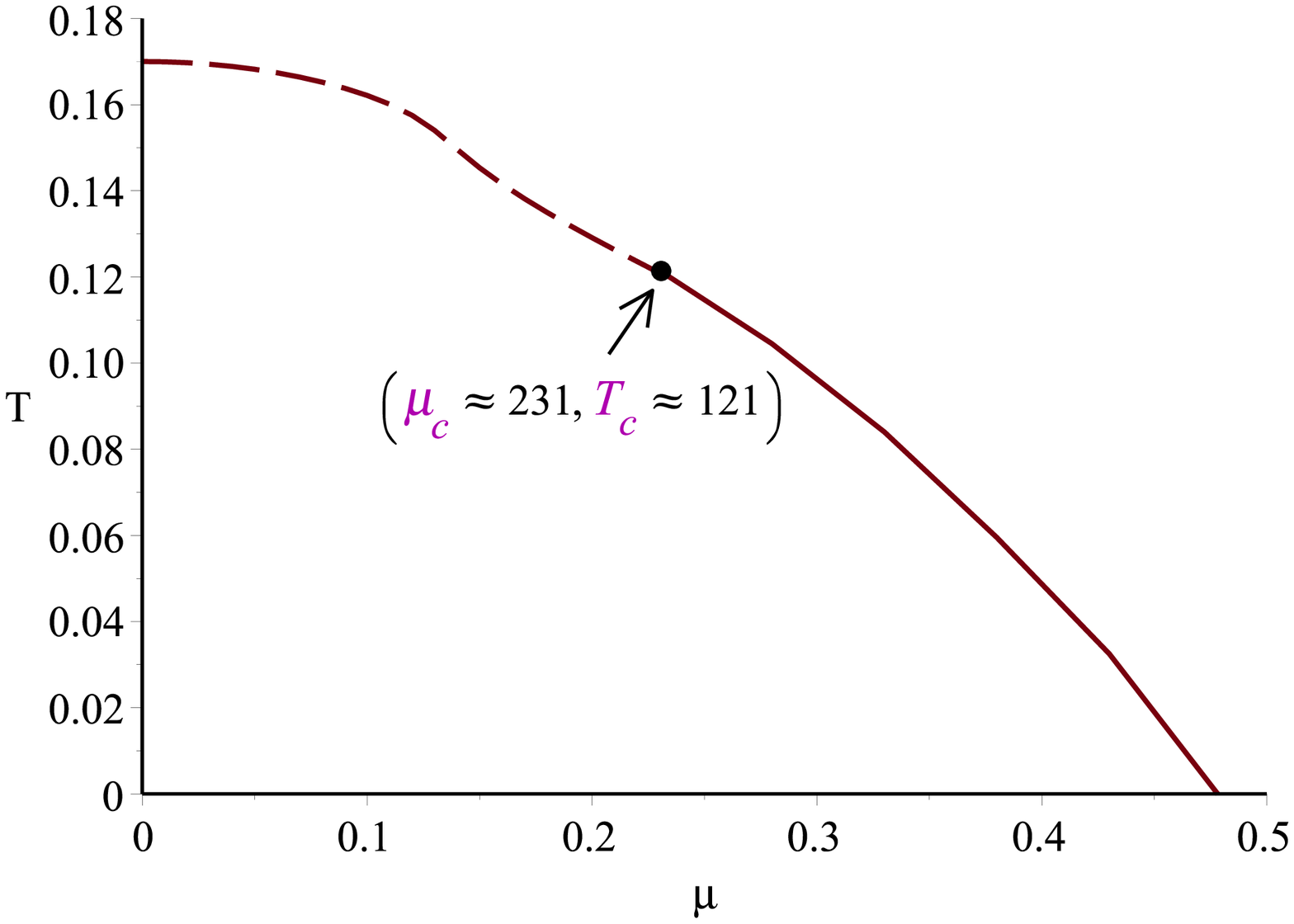}\vskip -0.05cm \hskip 0.15 cm \textbf{( a ) } \hskip 7.5 cm
\textbf{( b )}
\end{center}
\caption{The free energy v.s. temperature at different chemical potentials
$\mu$ is plotted in (a) and the phase diagram in $T$ and $\mu$ plane is
plotted in (b). For $0\leq\mu\leq\mu_{c}$, the free energies are single-valued
and smooth, the system undergoes a crossover. While for $\mu>\mu_{c}$, the
free energies become multi-valued and take swallow-tailed shapes with a
first-order phase transition happens at the self-crossing point. At $\mu
=\mu_{c}$, the free energy curve is single-valued but not smooth. A
second-order phase transition happens at the non-smooth point $\left(  \mu
_{c},T_{c}\right)  \simeq\left(  231MeV,121MeV \right)  $, which is the
critical point where the phase transition mildens to a crossover.}%
\label{phase}%
\end{figure}

The first law of thermodynamics in a grand canonical ensemble can be written
as,%
\begin{equation}
F=U-Ts-\mu\rho,
\end{equation}
where $U$ is the internal energy of the system and $F$ is the corresponding
free energy. Changes in the free energy of a system with constant volume are
given by%
\begin{equation}
dF=-sdT-\rho d\mu.
\end{equation}
At fixed values of the chemical potential $\mu$, the free energy can be
evaluated by the integral \cite{0812.0792,1301.0385}%
\begin{equation}
F=-\int sdT. \label{int F}%
\end{equation}
Directly integrating shows that the absolute value of the free energy goes to
infinity and needs to be regularized. However, since we only care about the
differences between the free energies, the absolute values of the free energy
are not important for our analysis. Thus we can simply regularize the free
energy by fixing the integration constant in the above integral (\ref{int F}).
Considering the vanishing chemical potential case, we set the free energy at
the quasi-transition temperature $T_{0}\approx170MeV$ to be zero. By
requesting $F\left(  T_{0}\right)  =0$ at $\mu=0$, we finally are able to
calculate the free energy as%
\begin{equation}
F=\int_{z_{H}}^{z_{H}\left(  T_{0}\right)  }s\dfrac{dT}{dz_{H}}dz_{H}.
\label{F}%
\end{equation}
The free energy $F$\ v.s. temperature $T$ and the phase diagram are plotted in
FIG. \ref{phase}.

As we expected, for $0\leq\mu\leq\mu_{c}$, the free energies are always
single-valued; while for $\mu>\mu_{c}$, the free energies become multi-valued
and take swallow-tailed shapes. A first-order phase transition happens at the
self-crossing point of each free energy curve with a fixed chemical potential.
At $\mu=\mu_{c}$, the free energy curve is continues but not smooth. A
second-order phase transition happens at the non-smooth point, which is the
critical point where the phase transition mildens to a crossover.

\subsection{Equations of State}

\begin{figure}[h]
\begin{center}
\includegraphics[
height=2in,
width=3in
]{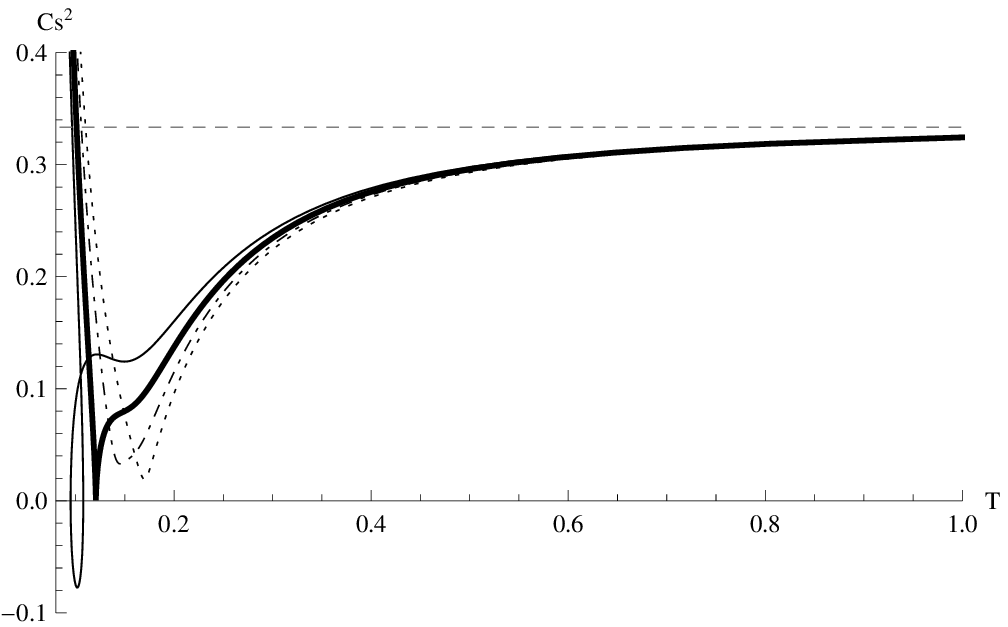}\hspace*{0.5cm} \includegraphics[
height=2in,
width=2.9in
]{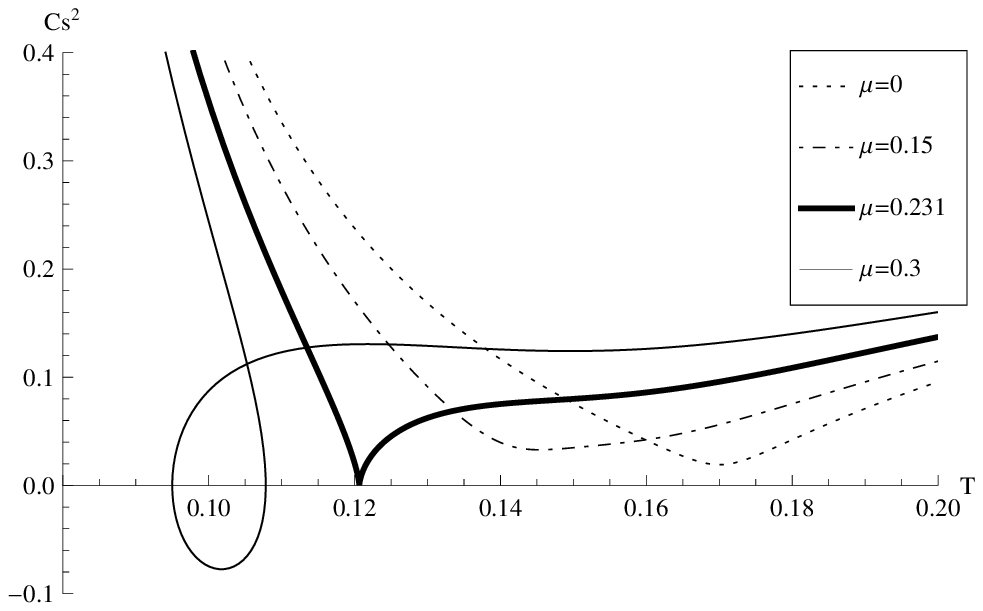}\vskip -0.05cm \hskip 0.15 cm \textbf{( a ) } \hskip 7.5 cm
\textbf{( b )}
\end{center}
\caption{The squared speed of sound v.s. temperature at different chemical
potentials $\mu=0,0.15,0.231,0.3GeV$. We enlarge a rectangle region in (a)
into (b) to see the detailed structure. For $0<\mu<\mu_{c}$, the speed of
sound behaves as a smooth crossover. At the critical point $\mu=\mu_{c}$, a
second order phase transition happens where $c_{s}^{2}$ goes to $0$ at the
critical temperature $T_{c}$. For $\mu>\mu_{c}$, the squared of speed of sound
undergoes a first order phase transition at the self-crossing point. At high
temperature, $c_{s}^{2}$ approaches the conformal limit $1/3$ as expected.}%
\label{cs-T}%
\end{figure}

FIG. \ref{cs-T} plots the squared of speed of sound $c_{s}^{2}$ v.s. the
temperature $T$ for different chemical potentials.

For $0<\mu<\mu_{c}$, the speed of sound behaves as a sharp but smooth
crossover. At the critical point $\mu=\mu_{c}$, a second order phase
transition happens where $c_{s}^{2}$ goes to $0$ at the critical temperature
$T_{c}$. For $\mu>\mu_{c}$, the squared of speed of sound becomes negative,
i.e. the speed of sound is imaginary, for a range of temperature. The
imaginary speed of sound indicates a Gregory-Laflamme instability
\cite{9301052,9404071}. This is related to the general version of Gubser-Mitra
conjecture \cite{0009126,0011127,0104071}, i.e. the dynamical stability of a
horizon is equivalent to the thermodynamic stability. In our system, the
negative specific heat implies thermodynamically unstable. While the imaginary
speed of sound implies the amplitude of the fixed momentum sound wave would
increase exponentially with time, reflecting the dynamical instability.
Roughly speaking, $C_{V}<0$ is equivalent to $c_{s}^{2}<0$ in our system. In
all the case, $c_{s}^{2}$ approaches the conformal limit $1/3$ at very high
temperature as expected.

\begin{figure}[h]
\begin{center}
\includegraphics[
height=2in,
width=3in
]{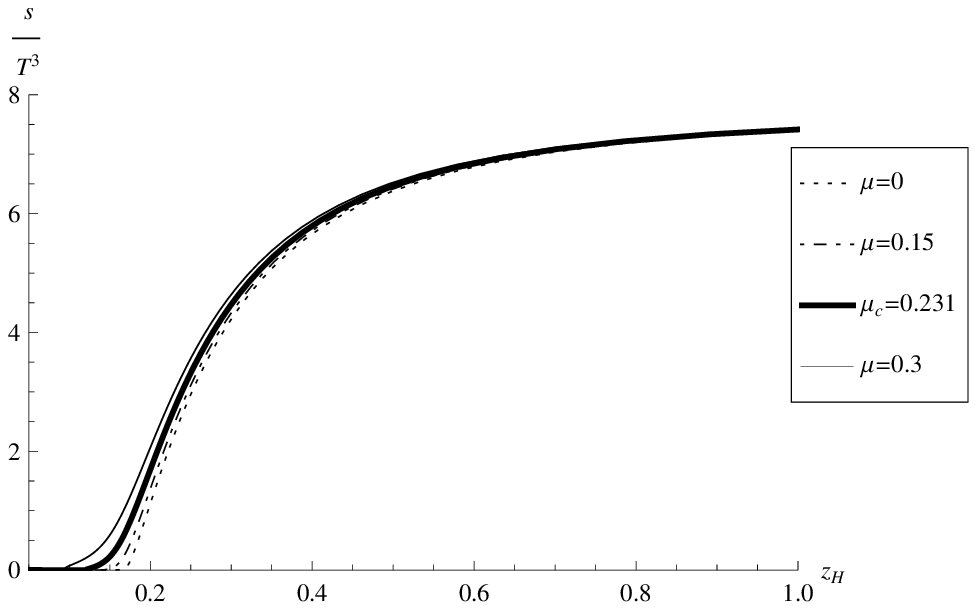}\hspace*{0.5cm} \includegraphics[
height=2in,
width=2.9in
]{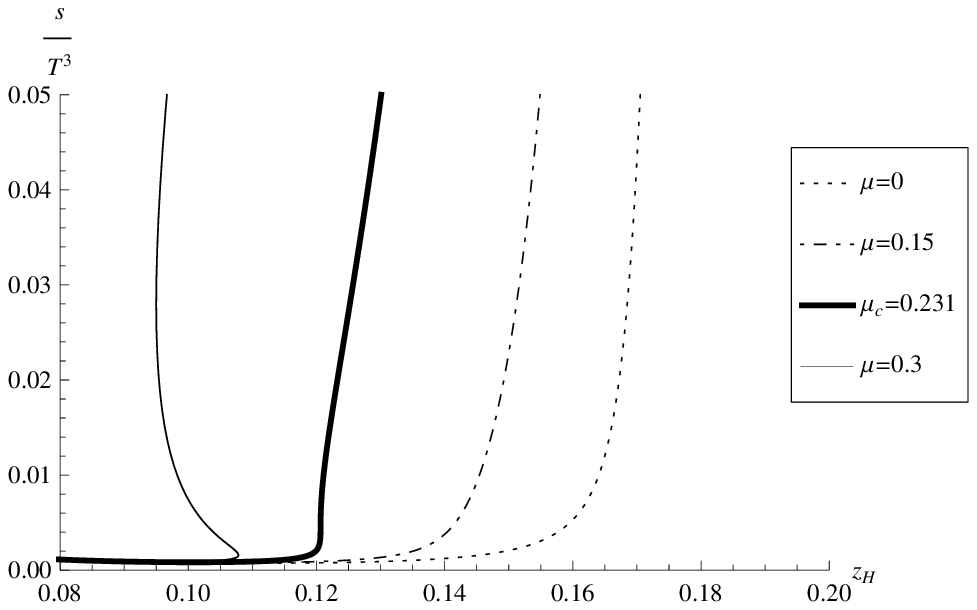}\vskip -0.05cm \hskip 0.15 cm \textbf{( a ) } \hskip 7.5 cm \textbf{(
b )}
\end{center}
\caption{The entropy v.s. temperature at different chemical potentials
$\mu=0,0.15,0.231,0.3GeV$. We enlarge a rectangle region in (a) into (b) to
see the detailed structure. For $0<\mu<\mu_{c}$, the entropy is single-valued
and there is no phase transition. For $\mu\geq\mu_{c}$, the entropy is
multi-valued for a region of temperature which indicates a phase transition
between high entropy and low entropy black holes.}%
\label{S}%
\end{figure}

We plot equations of state for entropy in FIG. \ref{S}. For $0<\mu<\mu_{c}$,
the entropy is single-valued and there is no phase transition. For $\mu\geq
\mu_{c}$, the entropy is multi-valued for a region of temperature which
indicates a phase transition between high entropy and low entropy black holes.
The similar phase behaviors have been discussed in \cite{0804.0434} for a
holographic QCD model with different values of parameters tuned by hand.

\begin{figure}[h]
\begin{center}
\includegraphics[
height=2in,
width=3in
]{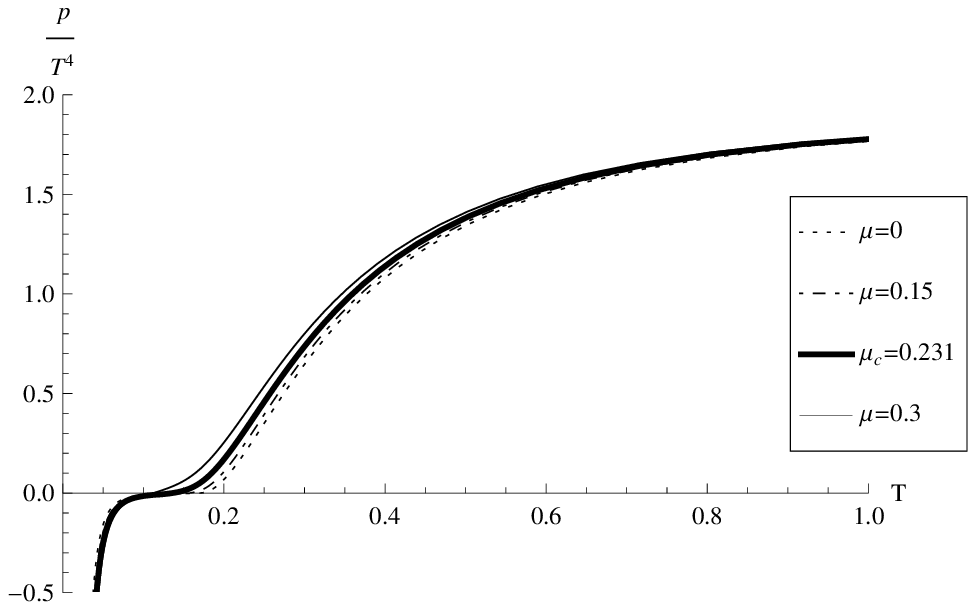}\hspace*{0.5cm} \includegraphics[
height=2in,
width=2.9in
]{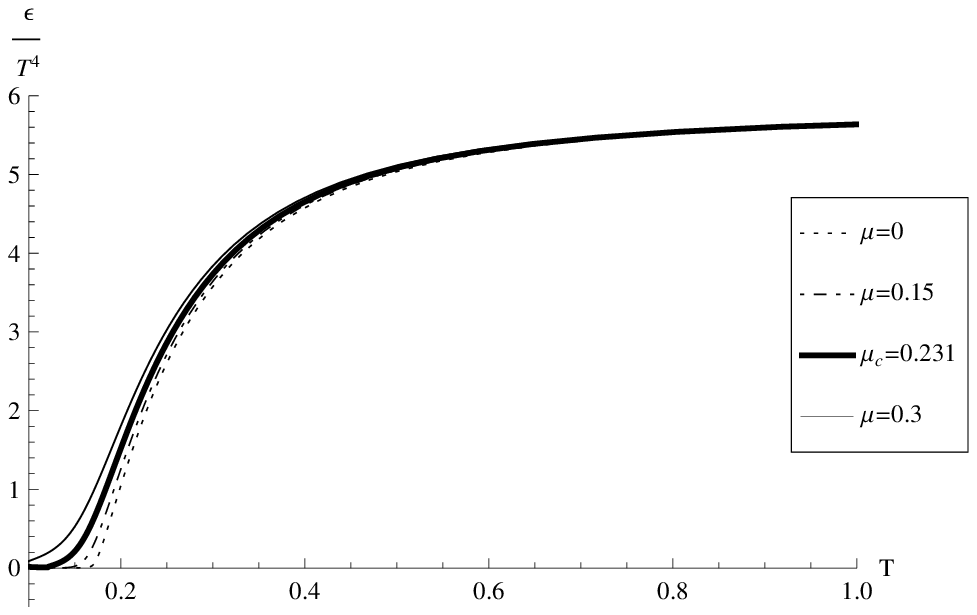}\vskip -0.05cm \hskip 0.15 cm \textbf{( a ) } \hskip 7.5 cm
\textbf{( b )}
\end{center}
\caption{The equations of state at different chemical potentials
$\mu=0,0.15,0.231,0.3GeV$. The pressure v.s. temperature is plotted in (a) and
the energy v.s. temperature is plotted in (b).}%
\label{EoS}%
\end{figure}

The pressure $p=-F$ and the energy $\epsilon=F+sT$ can be
calculated from the free energy and are plotted in FIG. \ref{EoS}. We see that both pressure and energy increases with the chemical potential, that pushes the
phase transition temperature to the smaller values for growing $\mu$. Our
results are consistent to the recent lattice results with finite chemical
potential \cite{1204.6710}.

\begin{figure}[h]
\begin{center}
\includegraphics[
height=2in,
width=3in
]{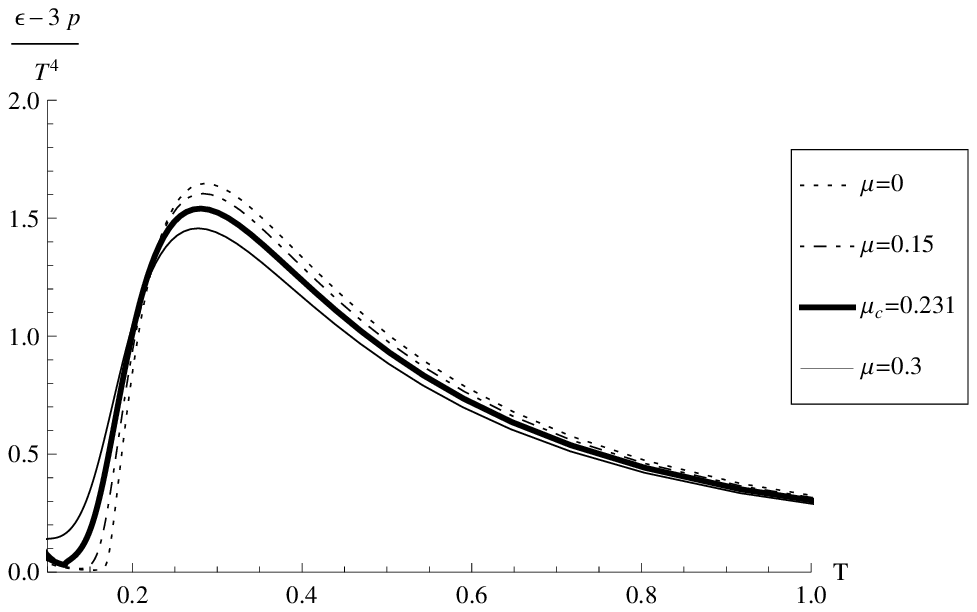}\hspace*{0.5cm} \includegraphics[
height=2in,
width=2.9in
]{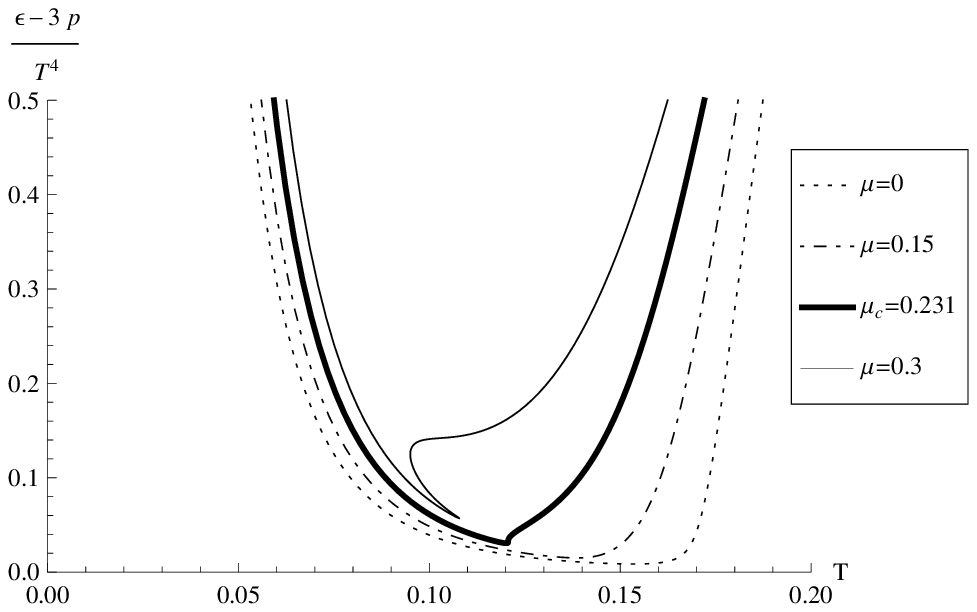}\vskip -0.05cm \hskip 0.15 cm \textbf{( a ) } \hskip 7.5 cm
\textbf{( b )}
\end{center}
\caption{The trace anomaly v.s. temperature at different chemical potentials
$\mu=0,0.15,0.231,0.3GeV$. We enlarge a rectangle region in (a) into (b) to
see the detailed structure. For $0<\mu<\mu_{c}$, the trace anomaly is single-valued
with finite slope through all the curve. For $\mu\geq\mu_{c}$, the slope of the trace anomaly becomes infinite at the certain temperature indicating a phase transition happened there.}%
\label{TA}%
\end{figure}

We finally plot the trace anomaly $\epsilon-3p$
v.s $T$ in FIG. \ref{TA}. With the growing chemical potential
$\mu$, the peak of trace anomaly decreases. From (b) in FIG. \ref{TA}, we clearly see that, for $0<\mu<\mu_{c}$, the trace anomaly is single-valued
with finite slope through all the curve. For $\mu\geq\mu_{c}$, the slope of the trace anomaly becomes infinite at the certain temperature indicating a phase transition happened there.

\setcounter{equation}{0}
\renewcommand{\theequation}{\arabic{section}.\arabic{equation}}

\section{Conclusion}

In this paper, we studied a Einstein-Maxwell-dilaton system with a dilaton
potential. We consistently solved the equations of motion of the system by the
potential reconstruction method. A family of analytic black hole solutions is
obtained. We then carefully studied the thermodynamic properties of the black
hole backgrounds. We computed the free energy to get the phase diagram of the
black hole backgrounds. In its dual holographic QCD theory, we are able to
realized the Regge trajectory of the vector mass spectrum by fixing the gauge
kinetic function. We then calculated the equations of state in our holographic
QCD model. We found that our dynamical model captures many properties in the
realistic QCD. The most remarkable feature of our model is that, by changing
the chemical potential, we are able to see the conversion from the phase
transition to a crossover dynamically. We identified the critical point in our
holographic QCD model and calculated its value with $\left(  \mu_{c}%
,T_{c}\right)  \simeq\left(  0.231GeV,0.121GeV \right) $. As the authors
knowledge, our model is the first holographic QCD model which could both
dynamically describe the transformation from the phase transition to the
crossover by changing the chemical potential and realize the linear Regge
trajectory for the meson spectrum.

There are many future directions are worth to be studied. For example, one can
introduce a open string in the black hole background and compute the linear
quark-antiquark potential and expectation value of Polyakov loop to
incorporate the confinement-deconfinement phase transition. One can also
compute the various transport coefficients like shear viscosity, bulk
viscosity and so on. It is also interesting to compute the critical exponents
of various physical quantities near the critical point. Some of these issues
are in progress.

\begin{acknowledgments}
We would like to thank Mei Huang, Xiao-Ning Wu for useful discussions. This
work is supported by the National Science Council (NSC 101-2112-M-009-005) and
National Center for Theoretical Science, Taiwan.
\end{acknowledgments}


\begin{thebibliography}{99}                                                                                               %


\bibitem {1009.4089}O.~Philipsen, \textquotedblleft Lattice QCD at non-zero
temperature and baryon density\textquotedblright\ [arXiv:1009.4089
[hep-lat]].




\bibitem {0306018}J.~Babington, J.~Erdmenger, N.~J.~Evans, Z.~Guralnik and
I.~Kirsch, \textquotedblleft Chiral symmetry breaking and pions in
nonsupersymmetric gauge / gravity duals,\textquotedblright\ Phys.\ Rev.\ D
\textbf{69}, 066007 (2004) [hep-th/0306018].




\bibitem {0311270}M.~Kruczenski, D.~Mateos, R.~C.~Myers and D.~J.~Winters,
\textquotedblleft Towards a holographic dual of large N(c)
QCD,\textquotedblright\ JHEP \textbf{0405}, 041 (2004)
[arXiv:hep-th/0311270].




\bibitem {0304032}M.~Kruczenski, D.~Mateos, R.~C.~Myers and D.~J.~Winters,
\textquotedblleft Meson spectroscopy in AdS / CFT with
flavor,\textquotedblright\ JHEP \textbf{0307}, 049 (2003)
[arXiv:hep-th/0304032].




\bibitem {0611099}S.~Kobayashi, D.~Mateos, S.~Matsuura, R.~C.~Myers and
R.~M.~Thomson, \textquotedblleft Holographic phase transitions at finite
baryon density,\textquotedblright\ JHEP \textbf{0702}, 016 (2007)
[arXiv:hep-th/0611099].




\bibitem {0412141}T.~Sakai and S.~Sugimoto, \textquotedblleft Low energy
hadron physics in holographic QCD,\textquotedblright%
\ Prog.\ Theor.\ Phys.\ \textbf{113}, 843 (2005) [arXiv:hep-th/0412141].




\bibitem {0507073}T.~Sakai and S.~Sugimoto, \textquotedblleft More on a
holographic dual of QCD,\textquotedblright\ Prog.\ Theor.\ Phys.\ \textbf{114}%
, 1083 (2005) [arXiv:hep-th/0507073].




\bibitem {0501128}J.~Erlich, E.~Katz, D.~T.~Son and M.~A.~Stephanov,
\textquotedblleft QCD and a holographic model of hadrons,\textquotedblright%
\ Phys.\ Rev.\ Lett.\ \textbf{95}, 261602 (2005) [arXiv:hep-ph/0501128].




\bibitem {0602229}A.~Karch, E.~Katz, D.~T.~Son and M.~A.~Stephanov,
\textquotedblleft Linear confinement and AdS/QCD,\textquotedblright%
\ Phys.\ Rev.\ D \textbf{74}, 015005 (2006) [arXiv:hep-ph/0602229].




\bibitem {0801.4383}B.~Batell and T.~Gherghetta, \textquotedblleft Dynamical
Soft-Wall AdS/QCD,\textquotedblright\ Phys.\ Rev.\ D \textbf{78}, 026002
(2008) [arXiv:0801.4383 [hep-ph]].




\bibitem {0806.3830}W.~de Paula, T.~Frederico, H.~Forkel and M.~Beyer,
\textquotedblleft Dynamical AdS/QCD with area-law confinement and linear Regge
trajectories,\textquotedblright\ Phys.\ Rev.\ D \textbf{79}, 075019 (2009)
[arXiv:0806.3830 [hep-ph]].




\bibitem {0804.0434}S.~S.~Gubser and A.~Nellore, \textquotedblleft Mimicking
the QCD equation of state with a dual black hole,\textquotedblright%
\ Phys.\ Rev.\ D \textbf{78}, 086007 (2008) [arXiv:0804.0434 [hep-th]].




\bibitem {1006.5461}U.~Gursoy, E.~Kiritsis, L.~Mazzanti, G.~Michalogiorgakis
and F.~Nitti, \textquotedblleft Improved Holographic QCD,\textquotedblright%
\ Lect.\ Notes Phys.\ \textbf{828}, 79 (2011) [arXiv:1006.5461 [hep-th]].




\bibitem {1012.1864}O.~DeWolfe, S.~S.~Gubser and C.~Rosen, \textquotedblleft A
holographic critical point,\textquotedblright\ Phys.\ Rev.\ D \textbf{83},
086005 (2011) [arXiv:1012.1864 [hep-th]].




\bibitem {1108.2029}O.~DeWolfe, S.~S.~Gubser and C.~Rosen, \textquotedblleft
Dynamic critical phenomena at a holographic critical point,\textquotedblright%
\ Phys.\ Rev.\ D \textbf{84}, 126014 (2011) [arXiv:1108.2029 [hep-th]].




\bibitem {1201.0820}R.~-G.~Cai, S.~He and D.~Li, \textquotedblleft A hQCD
model and its phase diagram in Einstein-Maxwell-Dilaton
system,\textquotedblright\ JHEP \textbf{1203}, 033 (2012) [arXiv:1201.0820
[hep-th]].




\bibitem {1103.5389}D.~Li, S.~He, M.~Huang and Q.~-S.~Yan, \textquotedblleft
Thermodynamics of deformed AdS$_{5}$ model with a positive/negative quadratic
correction in graviton-dilaton system,\textquotedblright\ JHEP \textbf{1109},
041 (2011) [arXiv:1103.5389 [hep-th]].




\bibitem {1209.4512}R.~-G.~Cai, S.~Chakrabortty, S.~He and L.~Li,
\textquotedblleft Some aspects of QGP phase in a hQCD model,\textquotedblright%
\ JHEP \textbf{1302} 068 (2013) [arXiv:1209.4512 [hep-th]].


\bibitem {1301.0385}Song He, Shang-Yu Wu, Yi Yang, Pei-Hung Yuan, "Phase
Structure in a Dynamical Soft-Wall Holographic QCD Model", JHEP \textbf{1304}
093 (2013) [arXiv:1301.0385 [hep-th]].



\bibitem {1111.4953}M.~Fromm, J.~Langelage, S.~Lottini and O.~Philipsen,
\textquotedblleft The QCD deconfinement transition for heavy quarks and all
baryon chemical potentials,\textquotedblright\ JHEP \textbf{1201}, 042 (2012)
[arXiv:1111.4953 [hep-lat]].


\bibitem {0912.2541}Pasi Huovinen and P\'{e}ter Petreczky, "QCD Equation of
State and Hadron Resonance Gas", Nucl.\ Phys.\ A\ 837 26-53 (2010)
[arXiv:0912.2541 [hep-ph]].

\bibitem {PDG2007}W. M. Yao et al. (Particle Data Group), J. Phys. G 33, 1
(2006) and 2007 partial update for the 2008 edition.

\bibitem {0710.0988}Mei Huang, Qi-Shu Yan, Yi Yang, "Confront Holographic QCD
with Regge Trajectories", Euro.\ Phys.\ J. C66, Issue 1 (2010) 187.
[arXiv:0710.0988 [hep-ph]].

\bibitem {0804.2731}Mei Huang, Qi-Shu Yan, Yi Yang, "Toward a more realistic
holographic QCD model", Prog.~Theor.~Phys. S~174, 334, (2008) [arXiv:0804.2731 [hep-ph]].

\bibitem {1403.1179}Jan M. Pawlowski, Fabian Rennecke, "Higher order quark-mesonic scattering processes and the phase structure of QCD", [arXiv:1403.1179 [hep-ph]].

\bibitem {1005.3508}S. Borsanyi et al., "Is there still any Tc mystery in
lattice QCD? Results with physical masses in the continuum limit III", JHEP
\textbf{1009} 073 (2010), [arXiv:1005.3508 [hep-lat]].



\bibitem {0812.0792}U.~Gursoy, E.~Kiritsis, L.~Mazzanti and F.~Nitti,
\textquotedblleft Holography and Thermodynamics of 5D
Dilaton-gravity,\textquotedblright\ JHEP \textbf{0905}, 033 (2009)
[arXiv:0812.0792 [hep-th]].




\bibitem {9301052}R.~Gregory and R.~Laflamme, \textquotedblleft Black strings
and p-branes are unstable,\textquotedblright\ Phys.\ Rev.\ Lett.\ \textbf{70},
2837 (1993) [arXiv:hep-th/9301052].




\bibitem {9404071}R.~Gregory and R.~Laflamme, \textquotedblleft The
Instability of charged black strings and p-branes,\textquotedblright%
\ Nucl.\ Phys.\ B \textbf{428}, 399 (1994) [arXiv:hep-th/9404071].




\bibitem {0009126}S.~S.~Gubser and I.~Mitra, \textquotedblleft Instability of
charged black holes in Anti-de Sitter space\textquotedblright%
\ [arXiv:hep-th/0009126].




\bibitem {0011127}S.~S.~Gubser and I.~Mitra, \textquotedblleft The Evolution
of unstable black holes in anti-de Sitter space,\textquotedblright\ JHEP
\textbf{0108}, 018 (2001) [arXiv:hep-th/0011127].




\bibitem {0104071}H.~S.~Reall, \textquotedblleft Classical and thermodynamic
stability of black branes,\textquotedblright\ Phys.\ Rev.\ D \textbf{64},
044005 (2001) [arXiv:hep-th/0104071].




\bibitem {1204.6710}S.~Borsanyi, G.~Endrodi, Z.~Fodor, S.~D.~Katz, S.~Krieg,
C.~Ratti and K.~K.~Szabo, \textquotedblleft QCD equation of state at nonzero
chemical potential: continuum results with physical quark masses at order
$mu^{2}$,\textquotedblright\ JHEP \textbf{1208}, 053 (2012) [arXiv:1204.6710
[hep-lat]].

\end{thebibliography}
\end{document}